\documentclass[aps,twocolumn,prl,preprintnumbers,amsmath,amssymb,superscriptaddress,nofootinbib]{revtex4-1}

 \usepackage{color}
 \usepackage{epsfig}
 \usepackage{ulem}

\definecolor{White}{rgb}{1,1,1}
\definecolor{Red}{rgb}{1,0.1,0}
\definecolor{LightYellow}{rgb}{1,1,.875}
\definecolor{SteelBlue}{rgb}{.273,.508,.703}
\definecolor{navy}{rgb}{0,0,.5}
\definecolor{LightCyan}{rgb}{.875,1,1}
\definecolor{DarkRed}{rgb}{.543,0,0}
\definecolor{HotPink}{rgb}{1,.41,.70}
\definecolor{ForestGreen}{rgb}{.13,.54,.13}
\definecolor{OliveDrab}{rgb}{.42,.55,.14}
\definecolor{MediumBlue}{rgb}{0,0,.80}
\definecolor{RoyalBlue}{rgb}{.25,.41,.88}
\definecolor{DeepSkyBlue}{rgb}{0,.746,1}
\definecolor{Brown}{rgb}{0.545,0.271,0.074}
\definecolor{Purple}{rgb}{0.637,0.285,0.641}

\def\bea{\begin{eqnarray}}
\def\eea{\end{eqnarray}}
\def\bec{\begin{center}}
\def\ec{\end{center}}

\def\beq{\begin{equation}}
\def\eeq{\end{equation}}

\newcommand\lsim{\mathrel{\rlap{\lower4pt\hbox{\hskip1pt$\sim$}}
    \raise1pt\hbox{$<$}}}
\newcommand\gsim{\mathrel{\rlap{\lower4pt\hbox{\hskip1pt$\sim$}}
    \raise1pt\hbox{$>$}}}
\def\bea{\begin{eqnarray}}
\def\eea{\end{eqnarray}}
\def\ba{\begin{array}}
\def\ea{\end{array}}
\def\bc{\begin{center}}
\def\ec{\end{center}}

\begin{document}

\title{Axionic Electroweak Baryogenesis}

\author{Kwang Sik Jeong} 
\email{ksjeong@pusan.ac.kr}
\affiliation{Department of Physics, Pusan National University, Busan 46241, Korea}
\author{Tae Hyun Jung} 
\email{thjung0720@ibs.re.kr}
\affiliation{Center for Theoretical Physics of the Universe, Institute for Basic Science (IBS), Daejeon, 34126, Korea}
\author{Chang Sub Shin} 
\email{csshin@ibs.re.kr}
\affiliation{Center for Theoretical Physics of the Universe, Institute for Basic Science (IBS), Daejeon, 34126, Korea}
%\date{}
\preprint{CTPU-18-13, PNUTP-18-A11}

\begin{abstract}

An axion can make the electroweak phase transition strongly first-order as required for electroweak
baryogenesis even if it is weakly coupled to the Higgs sector.
This is essentially because 
the axion periodicity naturally allows the structure of phase transition
 to be insensitive to the axion decay constant
 that determines the strength of axion interactions.
Furthermore, the axion can serve as a CP phase relevant to electroweak baryogenesis if one introduces
an effective axion coupling to the top quark Yukawa operator.
Then, for $f$ between about TeV and order $10$~TeV, the observed baryon 
asymmetry can be explained while avoiding current experimental constraints.
It will be possible to probe the axion window for baryogenesis in future lepton colliders and 
beam-dump experiments.

\end{abstract}

\pacs{}
\maketitle

Electroweak baryogenesis (EWBG) is one of the most attractive ways to explain the observed 
baryon asymmetry of the Universe~\cite{Kuzmin:1985mm, Shaposhnikov:1986jp, Shaposhnikov:1987tw}.
For successful EWBG, the Standard Model (SM) needs to be extended so that 
the electroweak phase transition (EWPT) is strongly first-order and an extra source 
of CP violation is present~\cite{Aoki:1999fi,Csikor:1998eu,Laine:1998jb,Gurtler:1997hr,Gavela:1993ts,Huet:1994jb,Gavela:1994dt}.
The connection to EWPT has led to consider new physics near the electroweak scale with 
significant couplings to the Higgs sector.   
However, it would then be difficult to reconcile EWBG with the observed properties of 
the SM-like Higgs boson and LHC null results for new particle searches so far while avoiding
the constraints from electric dipole moments (EDM) in association with CP violation relevant 
to baryogenesis.\footnote{ 
The limit on electron EDM has recently been improved by about one order of 
magnitude~\cite{Andreev:2018ayy}, constraining more severely the conventional EWGB scenarios. 
}

In this paper we present a scenario for EWBG where an axion $\phi$ weakly coupled to 
the Higgs sector induces a required strong first-order phase transition. 
The potential of the extended Higgs sector possesses 
a discrete shift symmetry, $\phi\to \phi+2\pi f$, 
and is written  
\begin{equation}
V = V(|H|^2, \sin\theta, \cos\theta), \label{Vgeneral}
\end{equation}
where $H$ is the Higgs doublet, and $\theta = \phi/f$.
Let us consider a simple model where $\phi$ couples to the Higgs squared operator\footnote{
	The axion-dependent potential terms can be generated nonperturbatively 
	in a controllable way, 
	for instance, see Ref.~\cite{Graham:2015cka}. 
	An example of a UV completion for the effective axion couplings to the Higgs 
   squared~\eqref{eq:ZTpotential} 
	and to the top quark Yukawa operator~\eqref{CPV} 
	is presented in the appendix. 
%	is presented in the appendix. 
}  
\begin{equation}
V_0 = \lambda |H|^4  + \mu^2|H|^2 
- M^2 \cos(\theta + \alpha)  |H|^2
- \Lambda^4 \cos\theta,
\label{eq:ZTpotential}
\end{equation}  
and so $\phi$ plays a crucial role in EWPT for $0<\mu^2 < M^2$.
%\footnote{
%
%The potential terms depending $\phi$ can be generated nonperturbatively, for instance, 
%if $\phi$ has an anomalous coupling to a hidden QCD,  
%and $H$ couples to additional vector-like lepton doublets and singlets via
%Yukawa interactions~\cite{Graham:2015cka}.
%Here the leptons are charged under the hidden QCD and have Majorana masses
%such that only the singlet leptons are relevant light degrees of freedom 
%at energy scales below the hidden confining scale.  
%gma
%}
The potential shape is determined by the constant phase $\alpha$ and the ratio between $M$ and $\Lambda$, 
but is insensitive to $f$ as the dependence on it arises only radiatively. 
This has an  important implication that the phase transition in the weak coupling limit, i.e.~at large $f$, is 
not SM-like differently from other scenarios. 
If one increases $f$, the field distance between potential minima gets large at the same time,
%along the axion direction, 
$|\Delta \phi| = f |\Delta \theta| \sim f \gg |\Delta H|$, 
and consequently the structure of phase transition remains approximately the same. 
This explains why EWBG is viable even if $\phi$ feebly couples to the Higgs sector. 

%This explains why EWBG is viable even if $\phi$ weakly couples to the Higgs sector
%differently from other scenarios. 
%Note also that a singlet-extended Higgs sector can generate a potential which allows
%a weakly coupled limit with a strong first-order phase transition as in our scenario.  
%However, it is only when one adds a proper dimension-six or higher operator since otherwise,
%the potential becomes unstable.  

One may wonder if a singlet scalar can play a similar role in EWPT as the axion.
Let us consider a simple extension with a singlet $s$ where
the high temperature potential develops a minimum at $s=H=0$, and a singlet-dependent Higgs 
mass squared is negative in a finite range of $s$. 
We assume that the Higgs mass squared has a minimum at $s=s_0$ and the electroweak vacuum
appears there.
A first-order phase transition is achieved if the potential has a proper profile in the region between
$s=0$ and $s=s_0$. 
It is possible to suppress the singlet coupling to $H$ while maintaining the shape of 
the potential $V(|H|^2,s/s_0)$, if one increases $s_0$ accordingly and imposes a specific 
hierarchical relation among singlet couplings.

How large can $f$ be without spoiling EWBG?  
It turns out that a strong first-order phase transition is possible for $f$ even larger than
$10^6$~GeV.
A more stringent bound comes from the observed baryon asymmetry.
To implement EWBG, one needs sizable CP violation at the phase interface, which can be provided 
for instance by an effective axion coupling to the top quark Yukawa operator. 
In such case, the axion itself corresponds to the CP phase and should have $f$ lower than order 
$10$~TeV to generate sufficient baryon asymmetry during phase transition.
This is because the wall width increases with $f$, which reduces baryon asymmetry for a given source
of CP violation. 
On the other hand, a lower bound on $f$ comes from the Higgs searches and results because $\phi$ mixes 
with the Higgs boson, and also from the EDM searches associated with the CP violation 
for baryogenesis.
Those constraints can be avoided for $f$ above about $1$~TeV. 
For $f$ in the multi-TeV range, it would be possible to probe our scheme by future experiments 
for axion-like particle searches.

For large $f$ above the electroweak scale, $H$ is much heavier than $\phi$.
Thus, for a qualitative understanding of a phase transition, one can approximately explore EWPT 
using the effective potential of a single light field, 
$\phi$, constructed by integrating out 
the heavy Higgs field via its equation of motion, 
i.e.~by replacing $H$ with the classical solution of $\partial V(H,\theta)/\partial H =0$~\cite{Jeong:2018jqe}:
\begin{equation}
\hat V = -\Lambda^4 \cos\theta + \Delta V(\theta), 
\label{eq:Vhat}
\end{equation} 
which is parameterized by
\begin{equation}
\Lambda, \quad \alpha, \quad
\epsilon \equiv \frac{\sqrt{2\lambda}\Lambda^2}{M^2},
\end{equation}  
for $\mu$ fixed by imposing the Higgs vacuum expectation value, $v=246$~GeV, at temperature 
$T=0$. 
Here we take $\alpha\geq 0$ without loss of generality. 
To simplify our discussion, we further approximate thermal corrections to include only those to the
Higgs mass squared.
Then, $\Delta V$ is non-vanishing in the interval of $\theta$ where $M^2\cos(\theta + \alpha)$ is larger
than the thermal-corrected Higgs mass squared. 
The effective potential of $\phi$ is enough to discern which region of the parameter space 
leads to a strong first-order phase transition.
After a qualitative description of EWPT, we provide precise results taking a two-field dimensional analysis
with full one-loop thermal corrections.

\begin{figure}[t] 
\begin{center}
\includegraphics[width=0.42\textwidth]{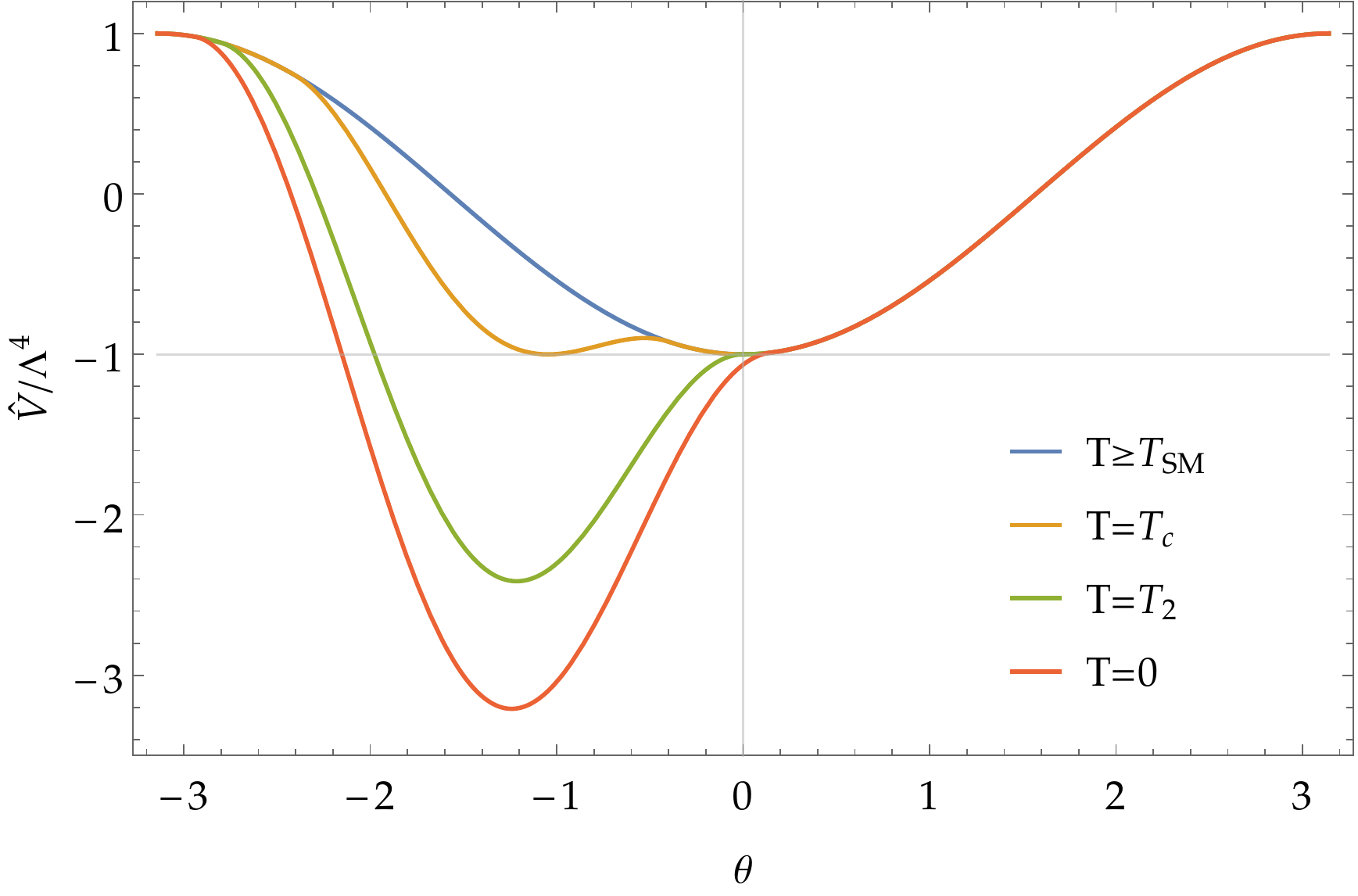} 
\includegraphics[width=0.44\textwidth]{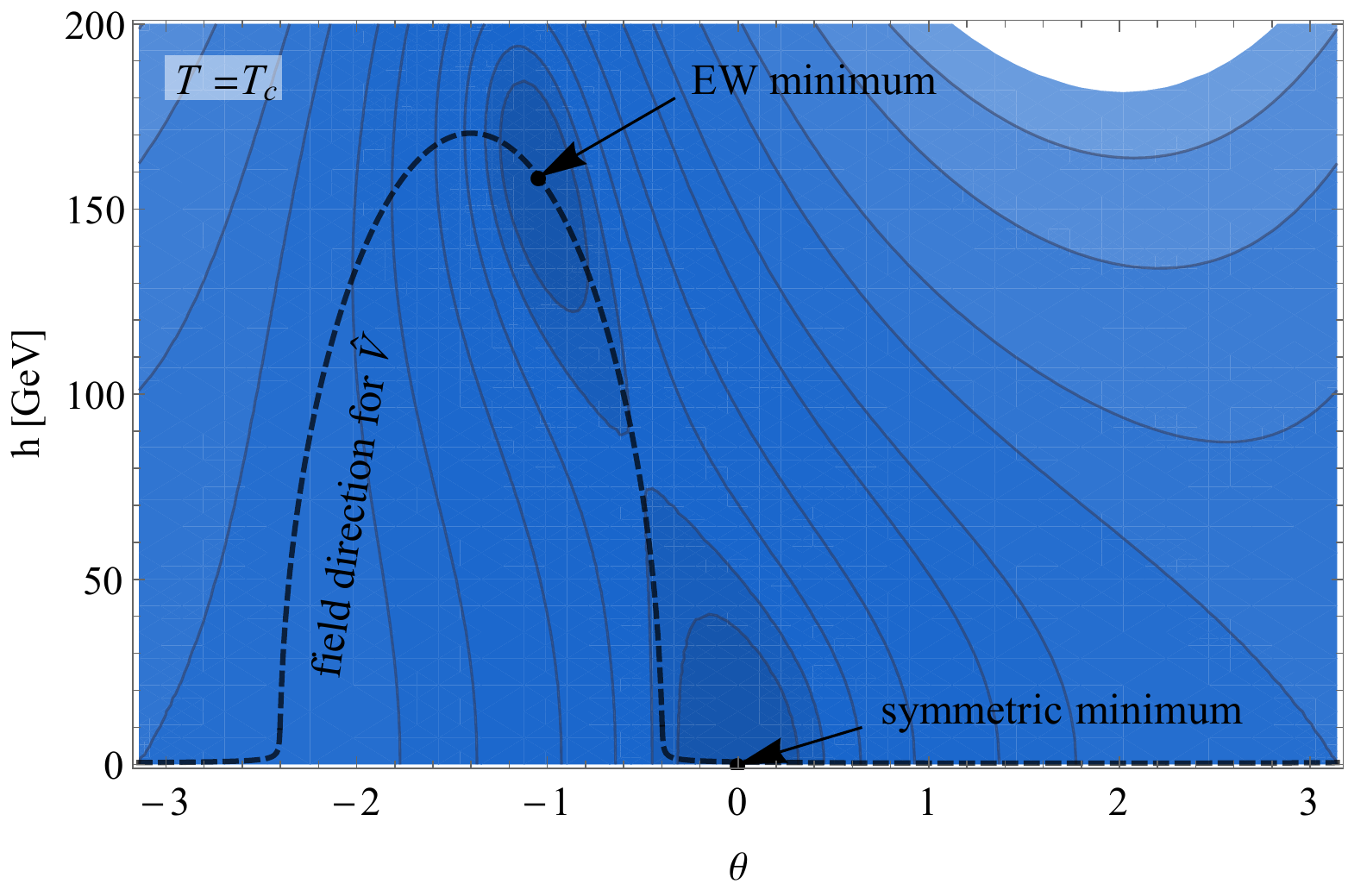} 
\end{center}
\caption{
Temperature-dependent potential 
for $\Lambda=80$~GeV, $\alpha=1.4$ and $\epsilon= 0.4$.
Here we have approximated thermal corrections to include only those to the Higgs mass squared. 
In the upper plot, which shows the effective potential $\hat V$, 
the blue curve is for $T\geq T_{\rm SM}$ with $T_{\rm SM}$ being the SM critical temperature.
The yellow, green and red curve are the potentials at $T=T_c,\,T_2$ and $0$, respectively.  
The lower plot is the potential at $T=T_c$ in the axion and Higgs field space, 
where brighter color represents a higher potential value and $\hat V$ describes the potential 
along the dashed line. 
 }
\label{fig:potential_FT}
\end{figure}

The upper plot of Fig.~\ref{fig:potential_FT} illustrates how a first-order phase transition is 
driven by the axion.
At high temperature (blue curve), a large thermal mass for $H$ leads to $\Delta V=0$, and 
the minimum of $\hat V$
is located at $\theta=0$.
An interval of $\theta$ where $H$ is tachyonic appears and increases as the Universe cools down,
and two degenerate minima are developed at $\theta=0$ and $\theta_c$ at 
the critical temperature $T_c$ (orange curve).
The electroweak minimum, $\theta=\theta_c$, becomes the true vacuum at a temperature below $T_c$
because it is deeper than the symmetric one, $\theta=0$.
In the lower plot, we show the contour plot of the potential at $T=T_c$ in the $(\theta,h)$ plane. 
The potential develops two minima separated by a barrier, and its behavior along the dashed line is 
described by $\hat V$.

\begin{figure}[t] 
	\begin{center}
		\includegraphics[width=0.36\textwidth]{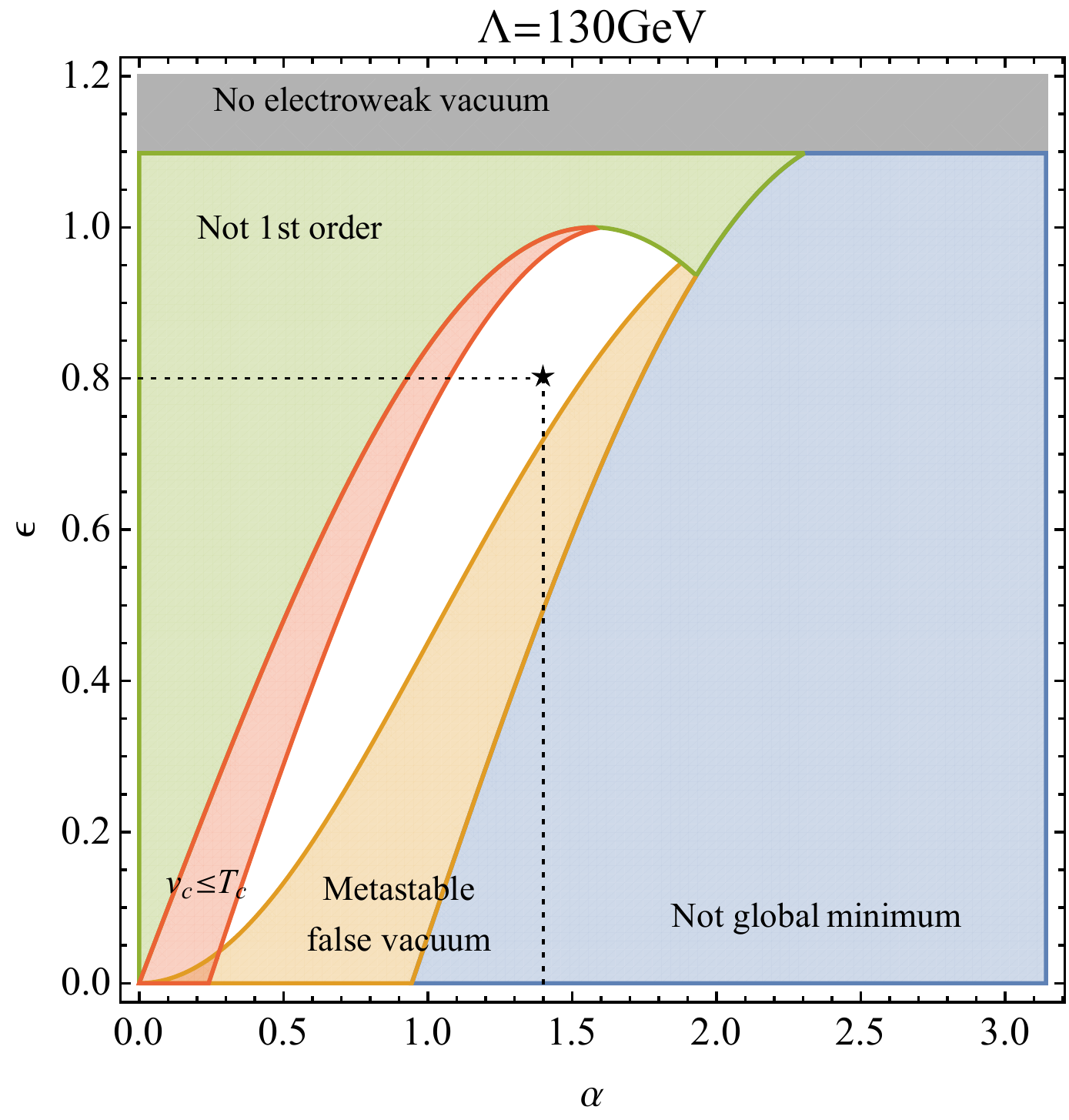} 
	\end{center}
	\caption{
	EWPT for $\Lambda=130$~GeV.   
	A strong first-order phase transition occurs in the white region.
	The symmetric minimum is deeper than the electroweak minimum in the blue region,
	while it is metastable in the orange region.
	The green and red regions are excluded because the transition is not first-order as in
	the SM, and is weak first-order, respectively.   
		}
	\label{fig:param}
\end{figure}

Let us explicitly examine the parameter region where EWPT is strongly first-order for
the case with $\Lambda=130$~GeV.
In the white region of Fig.~\ref{fig:param}, the axion-Higgs coupling induces a strong first-order phase 
transition, i.e.~$v_c/T_c>1$, where $v_c$ is the Higgs vacuum expectation value at $T_c$.
The blue region leads to no phase transition since the symmetric minimum is deeper than the
electroweak minimum, while the green region leads to the same situation as in the SM
phase transition. 
In the red region, a first-order phase transition takes place, but not strong.
The orange region is excluded because there is a potential barrier between two minima at $T=0$. 
If there remains a barrier at $T=0$, the vacuum transition rate is significantly suppressed
for $f$ above TeV, making the symmetric minimum metastable. 
For concreteness in the later discussion of baryogenesis, we choose a benchmark parameter point
\begin{equation}
\label{benchmark}
\Lambda=130\,{\rm GeV}, \quad
\alpha = 1.4, \quad \epsilon=0.8,  
\end{equation}
as marked by a star in the figure.
The benchmark case leads to a strong phase transition, $v_c/T_c\simeq 3$, with $T_c\simeq 68$~GeV 
and $T_2\simeq 54$~GeV.
Here $T_2$ denotes the temperature at which the barrier disappears.
One also finds that the axion mass reads $m_\phi \simeq 17\,{\rm GeV}\times (f/{\rm TeV})^{-1}$, and
the axion-Higgs mixing angle is given by $\delta \simeq 0.1 \times(f/{\rm TeV})^{-1}$,
which makes the axion detectable at colliders.
The estimated values are slightly changed depending on $f$ if one includes radiative and other thermal 
contributions in the potential.

Now we move on to analyzing the phase transition in detail and examining
the constraints on $f$.
Electroweak bubbles of the broken phase are nucleated at a temperature below $T_c$ and expand. 
The bubble nucleation rate per unit volume is roughly given by $T^4 e^{-S_3/T}$, and it exceeds
the Hubble rate when $S_3/T \approx 130$, which defines the nucleation temperature $T_n$.
Here $S_3$ is the Euclidean action of an O$(3)$-symmetric critical bubble~\cite{Coleman:1977py, Turok:1992jp}. 
Note that $S_3=0$ below $T_2$ because there is no potential barrier between two minima.

\begin{figure}[t] 
	\begin{center}
		\includegraphics[width=0.4\textwidth]{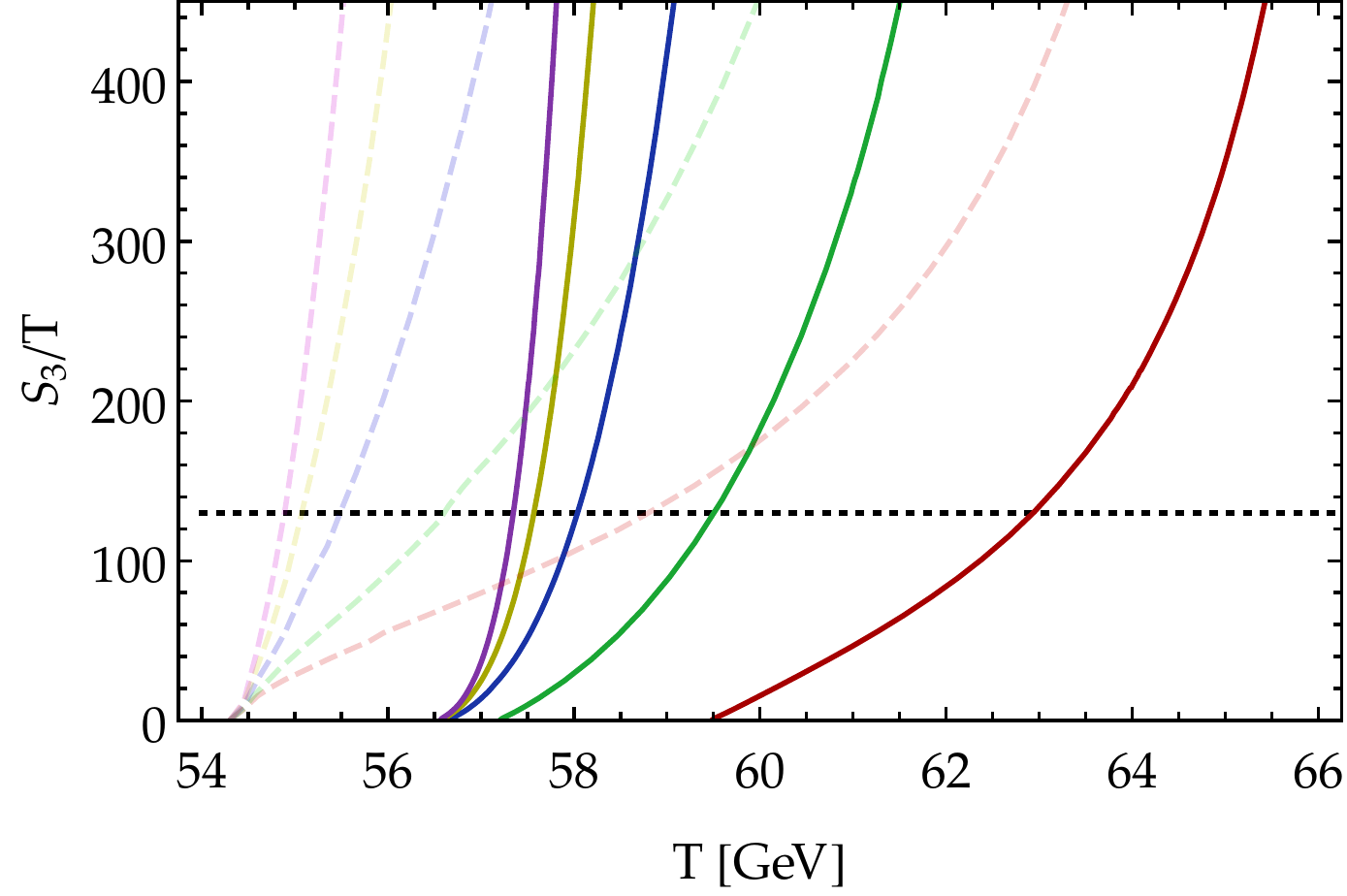} 
	\end{center}
	\caption{
	Bubble action for the benchmark case (\ref{benchmark}).
	The solid lines show $S_3/T$ for $f=0.2,\,0.5,\,1,\,1.5,\,2$~TeV from right to
	left, respectively, where full one-loop thermal corrections have been included.
	The dashed lines are obtained by taking only thermal corrections to the Higgs mass 
	squared.
	The dotted horizontal line is added to estimate $T_n$. 
     %	
	 %$S_3/T$ for the benchmark point \eqref{benchmark} 
	%	as functions of $T$ with various
	%	$f=200$, 500, 1000, 1500 and 2000 GeV from bottom to top.
	%	We include full one-loop thermal corrections to obtain the solid lines
	%	while dashed lines are obtained by using an approximate thermal correction
      %		($\Delta V_T = \frac{\gamma}{2}h^2T^2$).
	}
	\label{fig:S3overT}
\end{figure}

Fig.~\ref{fig:S3overT} shows $S_3/T$ as a function of $T$ for various values of $f$ 
in the benchmark case (\ref{benchmark}), where we have used the cosmoTransition python 
package to find a bubble solution through the overshooting-undershooting method
\cite{Wainwright:2011kj}.  
The dashed lines are obtained simply by the potential $V_0(h,\phi)$ including only the thermal corrections
to the Higgs mass squared, for which $T_2\simeq 54$~GeV regardless of $f$ as one can see in the figure.
On the other hand, the solid lines show the bubble action evaluated by taking the full one-loop 
potential of $h$ and $\phi$ at finite temperature~\cite{Quiros:1999jp}, $V(h,\phi)$, 
which depends radiatively on $f$.  
The nucleation temperature is determined by the intersection of each solid line with the dotted 
horizontal line, and it approaches to $T_2$ as $f$ increases. 
Here the barrier disappearing temperature is given by $T_2 \simeq 57$~GeV.  

It is interesting to notice that the Higgs contributions to the bubble action 
can be approximated by integrating out the Higgs field 
since the importance of its kinetic term gets suppressed as $f$ increases.
Consequently, 
$S_3$ is approximately given by $f^3$ times some function of $\Lambda$, $\alpha$, $\epsilon$ 
and $T$.
This feature can be understood because the action is written
\begin{equation}
S_3 = 4\pi f^3 \int x^2 dx \left( 
\frac{ h^{\prime 2}}{2 f^2} 
+ \frac{\theta^{\prime 2}}{2} 
+ V(h,\theta)
\right),
\end{equation}
showing that the Higgs contribution from $h^\prime$ is suppressed by $f^2$. 
Here $V$ includes radiative and thermal corrections, and the prime denotes a derivative with respect to 
$x\equiv r/f$ with $r$ being the radial distance from the center of the bubble.  
The approximate scaling behavior, $S_3\propto f^3$, reflects the fact that the effective potential
$\hat V(\phi)$  can describe well the phase transition
and the tunneling occurs mainly along the axion direction if $\phi$ is much lighter than $h$, 
that is, for large $f$.

Our scenario leads to $T_n$ rather close to $T_2$ compared to other scenarios, 
but the duration of phase transition,
$\Delta t_{\rm PT}$, 
is short enough to complete the first-order phase transition
before the barrier disappears.
In fact, a more relevant requirement is that  the time scale of generating baryon asymmetry, 
$\Delta t_{\rm BG}$, should satisfy  
\begin{equation}
\label{B-condition}
m^{-1}_\phi  < \Delta t_{\rm BG} < \Delta t_{\rm PT},
\end{equation}
where we have used that it takes time $\sim m^{-1}_\phi$ for $\phi$ to settle down to 
the true vacuum after quantum tunneling.
Since baryogenesis takes place outside bubbles via electroweak sphaleron processes, its time scale 
is the inverse sphaleron rate in the symmetric phase, 
$\Delta t_{\rm BG} \sim T^3/(\Gamma_{\rm sph}/V)$,
where $\Gamma_{\rm sph}/V\propto T^4$ is the sphaleron rate per unit 
volume~\cite{Arnold:1987mh, DOnofrio:2014rug}.
For large $f$, the bubble action shows the scaling behavior, $S_3 \propto f^3$, and 
$S_3/T$ is approximately linear in $T$ during the phase transition.
It thus follows that the duration of phase transition, which corresponds to the inverse of 
the time derivative of $S_3/T$~\cite{Megevand:2016lpr}, is roughly given by 
%$\Delta t_{\rm PT} \propto 1/(T^2 f^3)$ assuming radiation domination. 
$\Delta t_{\rm PT} \sim 10^{-2}\times (M_{\rm Pl}/T^2)({\rm TeV}/f)^3$,  
where $M_{\rm Pl}$ is the reduced Planck mass.
Note that the above condition \eqref{B-condition} puts an upper bound on $f$ because
both $m_\phi$ and $\Delta t_{\rm PT}$ are sensitive to $f$. 
For the benchmark case, baryogenesis requires $f$ below or around $10^6$~GeV.

To evaluate baryon asymmetry generated during the phase transition, one needs to specify 
the source of CP violation.
In our scenario, the axion mixes with the Higgs boson and its field value changes during phase transition.
Taking this into account, we introduce an axion 
coupling to the top quark Yukawa operator,
$e^{i\theta} \bar q_{L3} t_{R3} H$,
to get CP violation for baryogenesis.\footnote{
There are EWBG scenarios where the evolution of a misaligned axion field induces 
CP violation for baryogenesis~\cite{Kuzmin:1992up, Craig:2010au, Servant:2014bla}.
}

The effective top quark Yukawa coupling is then written
\begin{equation}
Y_t = y_t + x_t  e^{i(\theta + \beta)},
\label{CPV}
\end{equation}
for positive constants $y_t$ and $x_t$, taking an appropriate field redefinition. 
Thus, the axion plays the role of a CP phase, and the top quark acquires a mass whose phase 
varies across the wall depending on the bubble profile.
Note that the above axion coupling is subject to the EDM constraints if there is a relative
phase between the Higgs and axion couplings to the top quark, that is, between
$y_t+x_t e^{i\theta_0 + \beta}$ and $ix_t e^{i\theta_0 +\beta}$ with 
$\theta_0=\langle \phi\rangle/f$ at $T=0$ \cite{Baron:2013eja,Brod:2013cka,Fuyuto:2015ida}.
It is clear that the EDM constraints depend on $\beta$, but EWBG is insensitive 
to it for $x_t\ll 1$ because what matters in baryogenesis is the phase difference of 
the top quark mass across the wall.

\begin{figure}[t] 
	\begin{center}
		\includegraphics[width=0.36\textwidth]{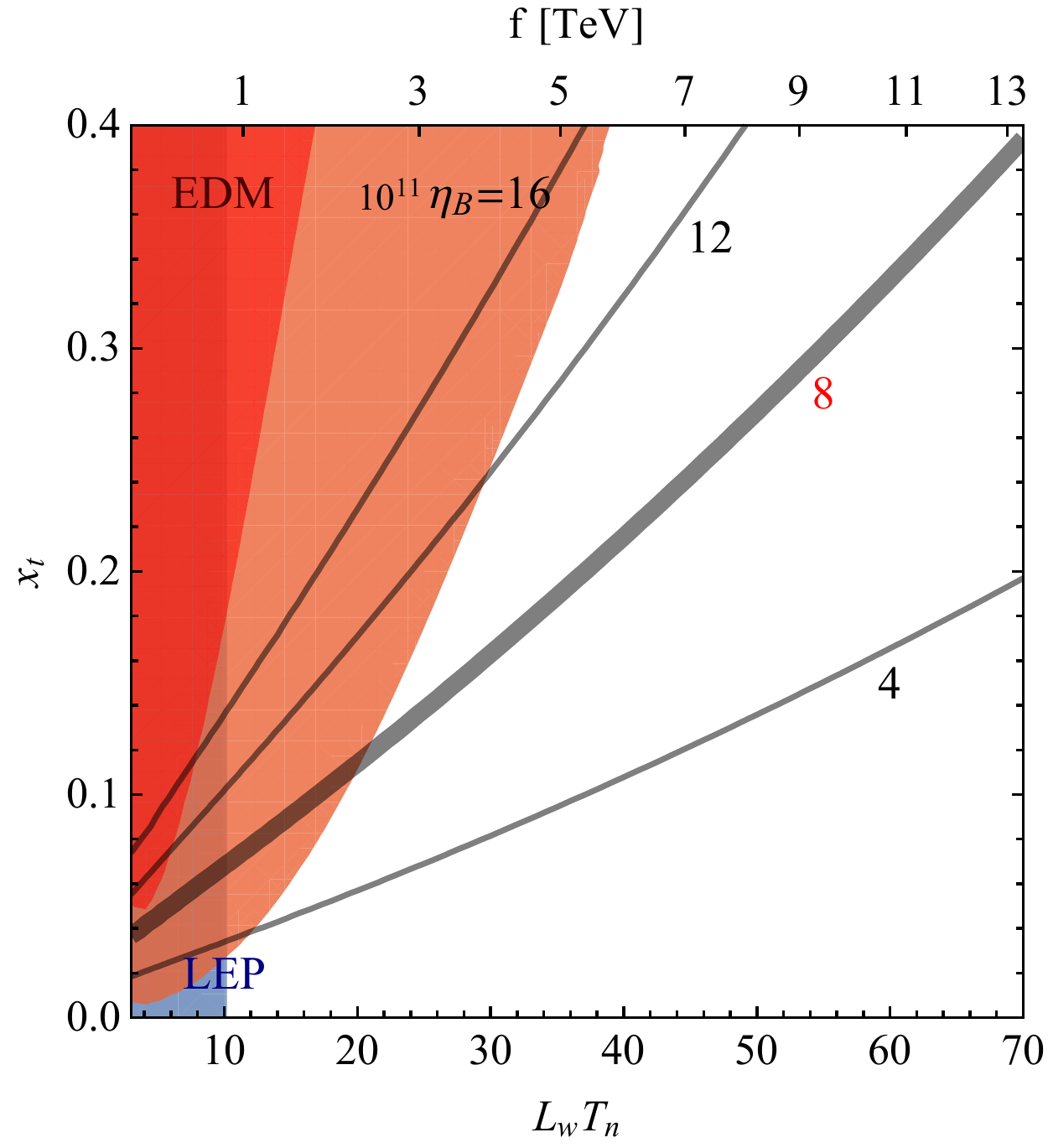} 
	\end{center}
	\caption{
	Baryon asymmetry for the benchmark case (\ref{benchmark})
	in the presence of an effective axion coupling to the top quark Yukawa operator.
	The red region is ruled out by the electron EDM constraint from ACME I \cite{Baron:2013eja}, which 
	requires 
	$|d_e|<8.7\times 10^{-29}\, e {\rm cm}$, while the light red region is ruled out by the recent measurement of ACME II,  $|d_e|<1.1\times 10^{-29}\, e{\rm cm}$~\cite{Andreev:2018ayy}. 
	The blue region is not 	compatible with the LEP results for Higgs searches.  
	}
	\label{fig:BA}
\end{figure}

For the benchmark case, we show in Fig.~\ref{fig:BA} the parameter region where the correct 
amount of baryon asymmetry is obtained while avoiding the electron EDM constraint.
Here we have taken $\beta \simeq -\theta_0$ so that the EDM contribution from the axion
is maximized.
The observed baryon asymmetry, $\eta_B = (8.2\,$--$\,9.4)\times 10^{-11}$, 
can be explained for $f$ below around $10$~TeV if one takes $x_t<0.3$. 
One can also see that the EDM constraint is important only for $f$ around or below TeV.
If one takes a different value of $\beta$, the EDM constraint gets weaker, and the generated
baryon asymmetry is slightly modified by a factor of order unity for $x_t$ above $0.1$.
It should be noted that $\eta_B$ is sensitive to the bubble dynamics, and our results are 
obtained under the assumptions summarized below.
A more careful study will be necessary to estimate the relic baryon asymmetry precisely.

The analysis assumes that the bubble profiles for $h$ and $\phi$ are approximated 
by a kink, $1-\tanh(z/L_w)$, where $z$ is the distance from the wall.
The wall width $L_w$ is numerically obtained by examining the 
$S_3$ critical bubble 
configuration, or equivalently $S_4$ 
as a time evolution of $S_3$ after bubble formation~\cite{Coleman:1977py},
and the relation between $f$ and $L_w$ is shown in Fig.~\ref{fig:BA}.
For a thick wall with $L_wT_n > 10$, the numerical computation suffers from instability when 
finding an inhomogeneous solution of the transport equations~\cite{Joyce:1994fu, Cline:2000nw,Bodeker:2004ws,Fromme:2006wx,Bruggisser:2017lhc}, and so we estimate the baryon asymmetry 
by performing an extrapolation, 
$\eta_B \propto x_t  (L_w/L_0)^{n_1 + n_2 \ln(L_w/L_0)}$, with constants $n_i$ and 
$L_0$.\footnote{ 
In this paper we shall focus on the case where baryon asymmetry is mostly produced away
from a bubble wall after diffusion. 
If the wall is much thicker, with $L_w T_n \gsim 100$, 
diffusion effects are less important and the dominant contribution to baryon asymmetry
is generated across the wall.
A successful realization of baryogenesis in such a limit has recently been 
discussed in Ref.~\cite{Jeong:2018jqe}.
% 
% 	In this paper, we only focus on the case where the baryon asymmetry is mostly
% 	generated away from the bubble wall after diffusion. 
% 	If the wall width is much larger ($L_w T_n \gsim 100$), then the diffusion effect is less important, 
% 	and the dominant contribution to the baryon asymmetry is generated across the wall. 
% 	A successful realization of baryogenesis in such a limit is discussed in Ref.\,\cite{Jeong:2018jqe}.
}
The baryon asymmetry also relies on the wall velocity $v_w$ at a stationary situation, which 
can be computed based on the fluctuation-dissipation theorem using that the pressure on the wall~\cite{Arnold:1993wc}
is determined by the potential difference between the broken and symmetric phases.
We find that the benchmark case leads to $v_w\simeq 0.07$.

Let us continue to discuss the experimental constraints on the axion properties and the 
future testability of our scenario. 
The LHC measurements~\cite{Aad:2015gba}, which require the Higgs signal strength 
relative to the SM prediction to be above $0.8$, constrain axion-Higgs mixing and Higgs decay into axions.
For the benchmark case, one needs $f$ above $340$~GeV, and the lower bound on $f$
would increase to $1.4$~TeV if the Higgs signal rate is measured at a sub-percent level
in future lepton colliders~\cite{Baer:2013cma}.  
A more stringent constraint comes from the Higgs searches at LEP because axion-Higgs mixing
allows the process, $e^+e^- \to Z \phi$~\cite{Flacke:2016szy}.
In the benchmark case, $\phi$ should be lighter than about $20$~GeV to avoid the LEP bound,
implying $f>850$~GeV.
For $2m_\mu<m_\phi \lesssim 5$~GeV, there is also an important constraint coming from 
rare meson decays, in particular from $B^+ \to K^+ \mu^+ \mu^-$.
To evade it, one needs
${\rm Br}(\phi \to \mu^+\mu^-) \times \sin^2 \delta\lesssim 6\times 10^{-7}$~\cite{Choi:2016luu},
implying roughly $ \sin\delta \lesssim 2\times 10^{-3}\times (m_\phi/{\rm GeV})$.
The constraint is thus strong for $f>3.4$~TeV in the benchmark case, but is avoidable
if $\phi$ couples to the QCD anomaly or a hidden sector. 
%to suppress ${\rm Br}(\phi \to \mu^+\mu^-)$.

% 
For the case that $\phi$ changes the phase of the top mass, there is a constraint
coming from cosmology and beam-dump experiments searching for axion-like-particles
if the axion has an anomalous coupling to photons, $\phi F\tilde F$ \cite{Alekhin:2015byh}.
Those constraints rule out $m_\phi$ smaller than $0.5$~GeV, from which we find the upper
bound on $f$ to be about $35$~TeV in the benchmark case.
It is thus weaker than the bound from the observed baryon asymmetry. 
In the presence of an axion-photon coupling, axion-Higgs mixing
generates extra EDMs~\cite{Choi:2016luu,Jung:2013hka}.
It should be emphasized that our model naturally avoids the EDM constraints 
for $f$ above about $3$~TeV because the baryon asymmetry is suppressed by 
$1/f^n$ with $n<1$, whereas the axion-induced EDM is strongly suppressed by $1/f^2$. 
As denoted in Fig.~\ref{fig:BA}, although the ACME II experiment has recently improved
the limit on electron EDM by one order of magnitude, % compared to the previous ones, 
our model for baryogenesis is still viable in a wide range of parameter space and does not
require any accidental cancellation.

Taking into account the constraints discussed above, we find that the viable range 
of the axion mass is approximately between $1.3$~GeV and $20$~GeV in the
benchmark case, which is obtained for the axion with $f$ between $0.85$~TeV and $13$~TeV. 
Interestingly, recasting the present LHC data to the search for axion-like particles sets
limits on $f$ for the indicated axion mass range~\cite{Knapen:2017ebd,Mariotti:2017vtv}.
High-luminosity LHC collisions are thus expected to probe our scenario.
The axion window can also be tested at future lepton colliders~\cite{Jaeckel:2015jla}.
%
%{\color{red} Recasting the present LHC data to search for axion like particles  (ALP)
%provides  interesting limits on the axion decay constant  for given axion mass ranges
%\cite{Knapen:2017ebd,Mariotti:2017vtv}. High-luminosity LHC collisions could give viable constraints 
%on our parameter space.}
%This parameter window also can be tested at future lepton colliders~\cite{Jaeckel:2015jla}.

Finally, we comment on how the viable parameter range changes 
when one considers a case different from the benchmark case.
If one moves away from the orange region in Fig.~\ref{fig:param}, $T_n$ and $T_2$ get higher, and thus 
the Higgs vacuum expectation value at $T_n$ decreases and the phase transition is weakened. 
As a result, the size of CP violation associated with the top quark is reduced during phase transition, 
lowering the upper bound on $f$.
On the other hand, if one approaches to the orange region, the upper bound on $f$ can be released, but 
instead, successful baryogenesis would require tuning of the parameters. 
The benchmark case has been chosen to implement baryogenesis without severe tuning while allowing large
$f$ as possible.
On the other hand, the lower bound on $m_\phi$ becomes weaker if one takes smaller $\epsilon$ since
the axion-Higgs mixing is proportional to it. 
 In this case, the LHC constraints on the Higgs properties get stronger than the LEP bound, restricting 
severely the parameter space to give, for instance, $m_\phi\lesssim {\cal O}(40)$~GeV for $\epsilon \sim 0.2$.

In this paper we have explored an axion-extended Higgs sector where EWPT is strengthened
enough to implement EWBG even when the axion is feebly coupled to the Higgs sector.
This is owing to the axion periodicity, which makes the profile of the scalar potential insensitive 
to the strength of axion interactions.
The structure of phase transition remains approximately the same under the change of
the value of $f$.  
It is also interesting that the axion itself
can serve as a CP phase for EWBG if one considers an effective axion
coupling to the top quark Yukawa operator.
In such case, axion-induced EWBG can account for the observed baryon asymmetry of the Universe
while evading the current experimental constraints if $f$ lies in the range between about TeV and order 
$10$~TeV. 
Note that roughly the axion mass reads $m_\phi \sim v^2/f$, and the axion-Higgs mixing angle 
is given by $\sin\delta \sim \alpha v/f$.
Therefore it will be possible to probe our scenario by future lepton colliders and beam-dump experiments.
\\ 
\\

\noindent{\bf Acknowledgments}
This work was supported by IBS under the project code, IBS-R018-D1 (THJ and CSS), 
and by the National Research Foundation of Korea (NRF) grant funded by the Korea
government (MSIP) (NRF-2018R1C1B6006061) (KSJ).
THJ and CSS thank to Eibun Senaha, Ryusuke Jinno, Fa Peng Huang and Mengchao Zhang 
for useful discussions.

\begin{appendix}

\section{Appendix}

Let us discuss how to generate effective axion couplings relevant to
axionic electroweak baryogenesis. 
% EWBG.
%
% to the Higgs squared
%operator (\ref{eq:ZTpotential}) and to the top Yukawa operator (\ref{CPV}). 
%  
The axion-dependent potential terms
\begin{eqnarray}
\label{DV}
%\Delta V =
 -\Lambda^4 \cos\theta -  M^2 \cos(\theta + \alpha) |H|^2
\end{eqnarray}
have been considered to make
electroweak phase transition
% EWPT 
strongly first-order, and the axion coupling to the 
top quark Yukawa operator
\begin{eqnarray}
\label{DL}
%\Delta {\cal L} = 
x_t e^{i\theta} H q_3 t^c % + {\rm h.c.}
\end{eqnarray}
has been added as an example of CP violation for baryogenesis.
Here we have used two-component Weyl spinors.
It is obvious that the above interactions break shift symmetry, 
$\theta \to \theta + ({\rm constant})$, 
but respect the discrete shift symmetry, $\theta \to \theta +2\pi$, 
as required by the axion periodicity. 
 
\begin{figure}[t] 
	%\begin{center}
		\includegraphics[height=0.14\textheight]{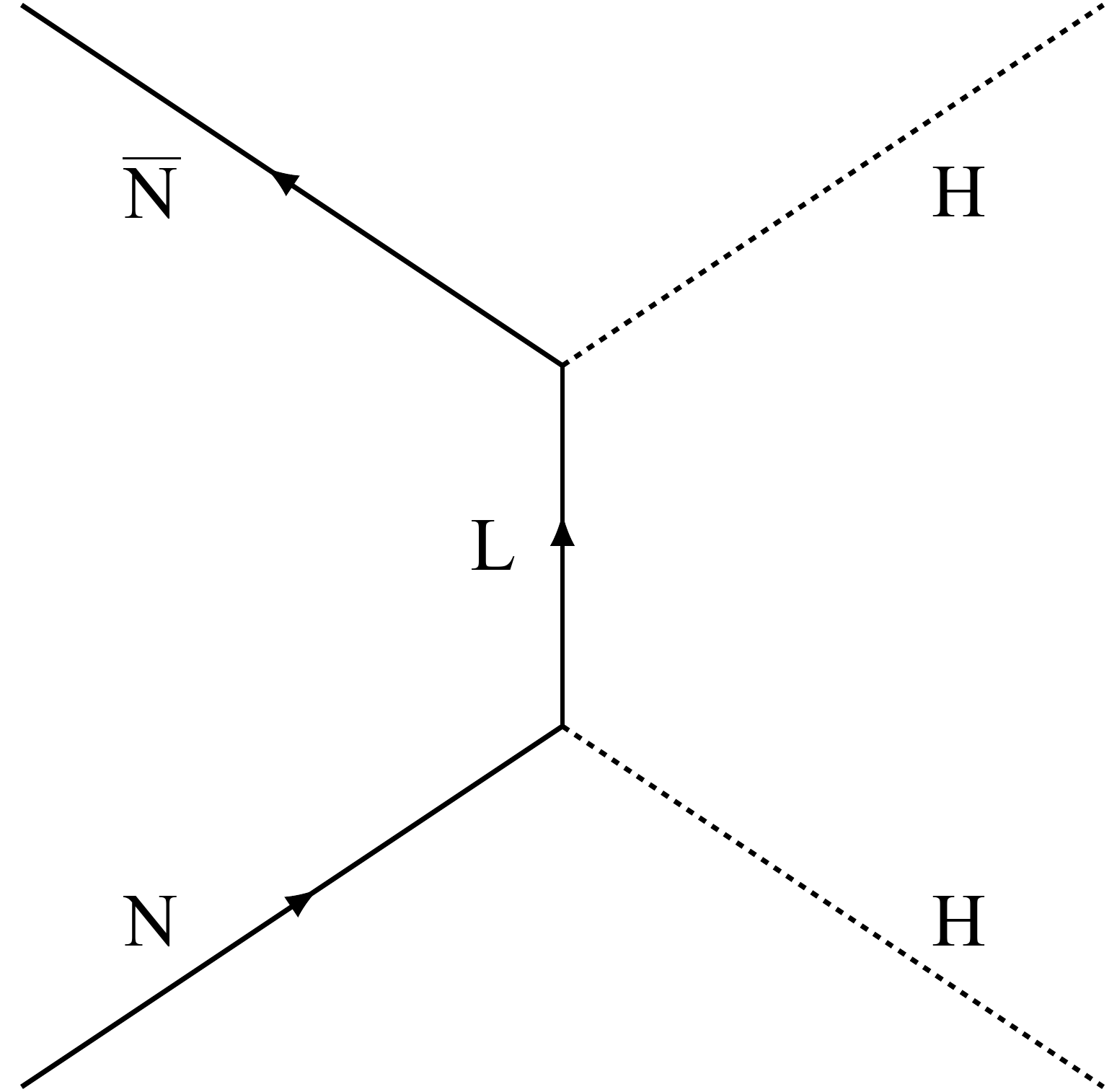} ~~~~~
		\includegraphics[height=0.14\textheight]{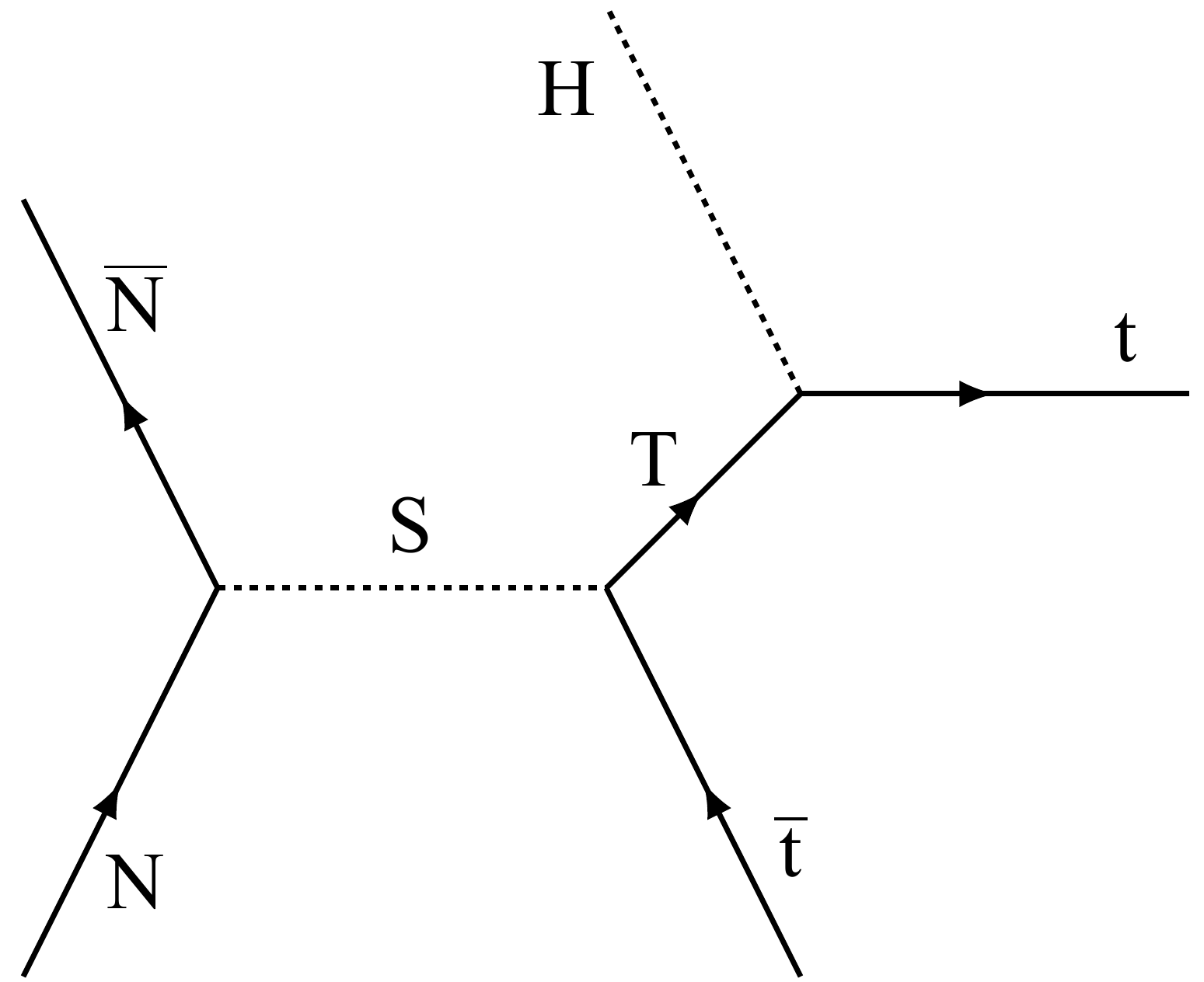} 
\vspace{0.5cm}
	%\end{center}
\\
		\includegraphics[height=0.116\textheight]{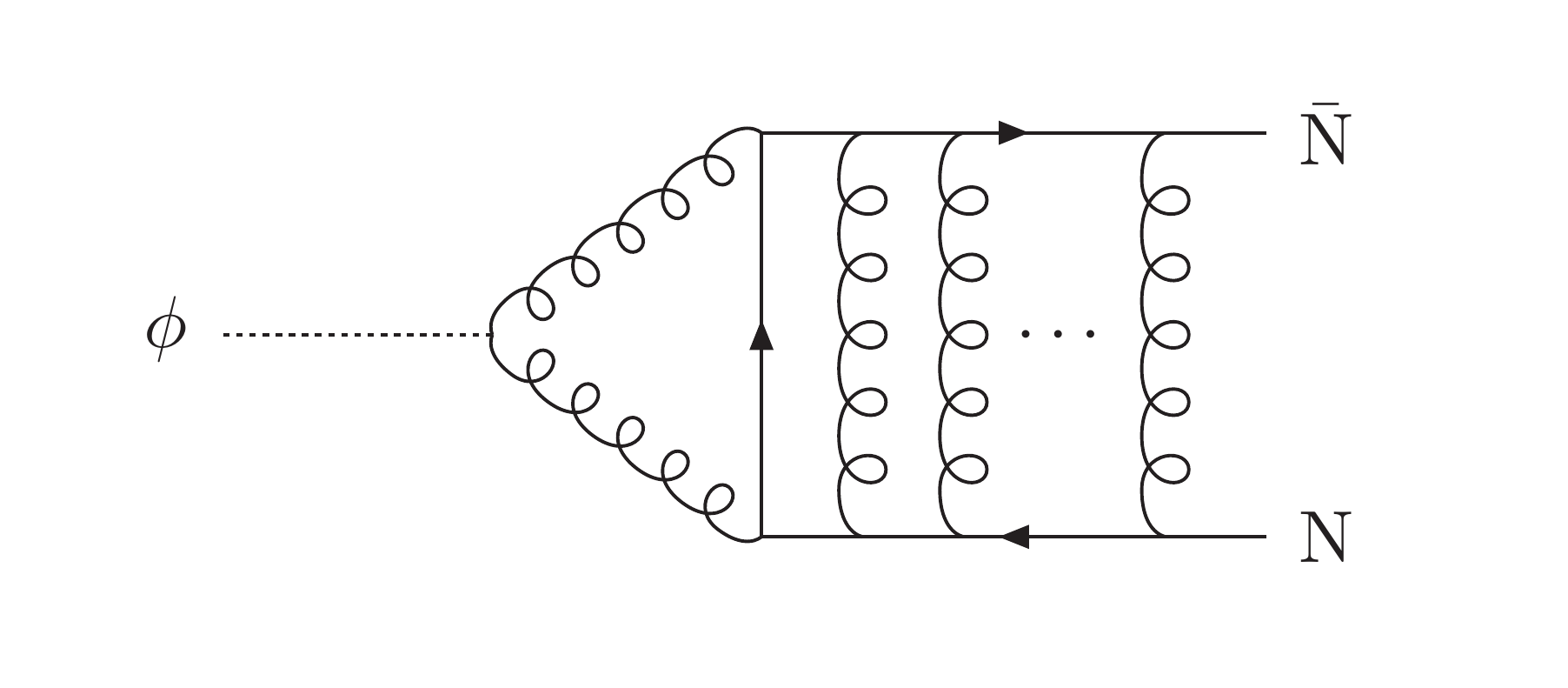} 
	\caption{ 
Feynman diagrams for effective axion couplings. 
After hidden quarks $N+\bar N$ condensate, the axion 
mixes with the hidden meson (lower panel) due to the anomalous axion coupling to hidden QCD,
 and consequently couples to 
the Higgs squared operator (upper left panel) and to the top quark
Yukawa operator (upper right panel).
	}
	\label{fig:diagram}
\end{figure}

It is possible to generate the axion interactions in a controllable way if one considers nonperturbative
effects breaking the axion shift symmetry.  
As an explicit example, we consider additional vector-like lepton doublets $L+L^c$ and 
singlets $N+N^c$ which are charged
under a hidden QCD confining at $\Lambda_c$, and have interactions 
\begin{eqnarray}
%{\cal L} \ni
y H L N^c + y^\prime H^\dagger L^c N + m_L LL^c + m_N NN^c.
\end{eqnarray}
Note that the above interactions preserve the axion shift symmetry.
For $m_N < \Lambda_c < m_L$, the heavy lepton doublets are integrated out to induce
an effective mass term for $N+N^c$,
%a low energy effective theory can be obtained by integrating 
%out the heavy lepton doublets:
\begin{eqnarray}
%{\cal L}_{\rm eff} \ni
\left( m_N + \frac{yy^\prime }{m_L} |H|^2 \right) NN^c. 
\end{eqnarray} 
The axion shift symmetry can be anomalous under hidden QCD if it is linearly realized and spontaneously broken
while making some hidden quarks massive.  
Then, the axion has a coupling
\begin{eqnarray}
\label{axion-FF}
\frac{1}{32\pi^2} \theta F \tilde F,
\end{eqnarray} 
where $F$ is the field strength of hidden QCD, and $\tilde F$ is its dual. 
The light hidden quarks $N+N^c$ condensate at $\Lambda_c$, breaking the associated 
chiral symmetry.
As a result, there arise potential terms, 
\begin{eqnarray}
-\Lambda^4_c \cos(\theta_N + \theta) -
\left( m_N + \frac{yy^\prime}{m_L}  |H|^2 \right)\Lambda^3_c \cos\theta_N,
\end{eqnarray}
at energy scales below $\Lambda_c$.
Here $\theta_N \equiv \arg(NN^c)$ denotes the meson field. 
The first term induces axion-meson mixing, and thus leads to
\begin{eqnarray}
\Delta V_{\rm eff} \approx
-\left( m_N + \frac{yy^\prime}{m_L} |H|^2 \right)\Lambda^3_c \cos\theta,
\end{eqnarray}
after integrating out the heavy meson.  
%It is worth noting that the situation is similar to the QCD axion that solves the strong CP problem.
%One can see that the axion couples to the Higgs squared operator as a result of meson-axion mixing,
It should be noted that the effective axion coupling to the Higgs squared operator
is controllable because it vanishes in the limit that the hidden QCD gauge coupling goes to zero. 
The Feynman diagrams for the effective axion coupling are presented in Fig.~\ref{fig:diagram}. 

% \begin{figure}[t] 
%	%\begin{center}
%		\includegraphics[height=0.08\textheight]{axion_meson_mixing.pdf} 	%\end{center}
%	\caption{Feynman diagram for the mixing between the axion and the hidden sector meson. 
%	The axion anomalous coupling to hidden QCD breaks the axion shift symmetry explicitly, and generates
%	axion couplings to the Higgs squared operator (left) and the top Yukawa operator (right) of Fig.~\ref{fig:diagram}
%	via axion-meson mixing after a condensation of $N+N^c$. {\color{red} adding the field id for each vertex}
%	}
%	\label{fig:mixing}
%\end{figure} 

%Let us continue to examine how to obtain the axion coupling to the top quark Yukawa operator. 

The axion coupling to the top Yukawa operator can be generated in a similar way.
Let us introduce lepton singlets $N+N^c$, vector-like top quark partners $T+T^c$, 
and a real scalar $S$
with masses, $m_N$, $m_T$, and $m_S$, respectively.
They have interactions
\begin{eqnarray}
\lambda SNN^c + \lambda^\prime S T t^c + \kappa H\bar q_3 T^c,
%\nonumber \\
%&&
%+\, m_N NN^c + m_T TT^c  - \frac{1}{2}m^2_S S^2,
\end{eqnarray}
which preserve the axion shift symmetry. 
For $m_T,\,m_S \gg m_N$, it is straightforward to see that the low energy effective theory 
of $N+N^c$, which are colored under hidden QCD, is described by
\begin{eqnarray}
%{\cal L}_{\rm eff} \ni
\left( m_N + \frac{2\kappa \lambda \lambda^\prime}{m^2_Sm_T}   
Hq_3 t^c \right) NN^c.
\end{eqnarray}
%which is obtained by integrating out the heavy fields.  
For $m_N < \Lambda_c$, the hidden quarks condensate to give
\begin{eqnarray}
\hspace{-5mm}
\Lambda^4_c \cos(\theta_N + \theta) +
\left( m_N + \frac{2\kappa \lambda \lambda^\prime}{m^2_Sm_T}   
Hq_3 t^c \right)  \Lambda^3_c \cos\theta_N,
\end{eqnarray}
where the first term arises from the axion anomalous coupling to hidden QCD. 
Thus, integrating out the heavy meson $\theta_N$ leads to
\begin{eqnarray}
\Delta {\cal L}_{\rm eff} \approx 
\left( m_N + \frac{2\kappa \lambda \lambda^\prime}{m^2_Sm_T}   
Hq_3 t^c \right)  \Lambda^3_c \cos\theta,
\end{eqnarray}
showing that the axion couples to the top Yukawa operator, which is essentially due to  
axion-meson mixing induced by the axion anomalous coupling to hidden QCD that breaks
the axion shift symmetry.
Fig.~\ref{fig:diagram} illustrates how the effective axion coupling to the top Yukawa
operator arises.

%Let us summarize our discussion.
The mass parameters in the potential (\ref{DV}) and the coupling in the interaction
(\ref{DL}) are given in terms of the couplings of high energy theory,
\begin{eqnarray}
\Lambda^4 &=& m_N \Lambda^3_c,
\nonumber \\
M^2 &=& \frac{yy^\prime \Lambda^3_c}{m_L},
\nonumber \\
x_t &=& \frac{2\kappa\lambda\lambda^\prime \Lambda^3_c}{m^2_Sm_T},
\end{eqnarray}
and the constant phase $\alpha$ in ($\ref{DV}$) is determined
by $\arg(yy^\prime) - \arg(m_N m_L)$.
Hence the benchmark case with $\Lambda=130$~GeV, $M=100$~GeV and $x_t={\cal O}(0.1)$
is obtained for instance by taking $\Lambda_c\sim$~TeV, 
$m_N\sim0.3$~GeV, $m_L\sim100$~TeV, and $m_S\sim m_T\sim 2$~TeV for Yukawa 
couplings of order unity.

\end{appendix}

\end{document}